\documentclass[aps,prl,twocolumn,superscriptaddress]{revtex4-1}

\usepackage{graphicx}

	\newcommand{\hatgam}{\hat{\gamma}}
	\newcommand{\hnab}{\hat{\nabla}}
	\newcommand{\hLam}{\hat{\Lambda}}

\begin{document}


\title{Non-axisymmetric oscillations of rapidly rotating relativistic stars
by conformal flatness approximation}


\author{Shin'ichirou Yoshida}
\email[]{yoshida@ea.c.u-tokyo.ac.jp}
\affiliation{Department of Earth Science and Astronomy, Graduate School
of Arts and Science, The University of Tokyo\\
Komaba, Meguro-ku, Tokyo 153-8902, Japan}


\date{\today}

\begin{abstract}
We present a new numerical code to compute non-axisymmetric eigenmodes of
rapidly rotating relativistic stars by adopting spatially conformally flat approximation
of general relativity. The approximation suppresses the radiative degree of
freedom of relativistic gravity and the field equations are cast into
a set of elliptic equations.
The code is tested against the low-order f- and p-modes of slowly rotating stars for which a good agreement is observed in frequencies computed by our new code
and those computed by the full theory. Entire sequences of the low order counter-rotating f-modes are computed, which are susceptible to an instability driven by gravitational radiation.
\end{abstract}


\maketitle

\section{Introduction}
%
Computing characteristic oscillations of rapidly rotating stars in general
relativity (GR) is much harder than in Newtonian
theory where it is already a difficult task because of deformations
of equilibria from a spherical figure.
Relativistic gravity 
is expressed by a spacetime metric tensor
which has its own dynamical degrees of freedom, compared to a single scalar potential in Newtonian gravity.
These dynamical degrees of freedom are encoded in the theory in such an intricate
way that they may be separated from non-dynamical ones only in cases with
a high degree of symmetry.  
The equilibrium far from a sphere
raises another important issue of how to impose 
boundary conditions for gravitational radiation. 
Free oscillations of a star in GR are characterized by a condition of 
gravitational waves that propagate "purely outward"
at the infinity. For rapidly rotating stars, there have been so far no prescription
to impose this condition.

Up to now the only linear eigenmode problem of rapidly rotating stars solved in full GR is that of the neutral mode when the eigenfrequency vanishes (\citet{Stergioulas_Friedman1998}, hereafter SF98).
For rotating stars in post-Newtonian theory \citet{Cutler_Lindblom1992} performed eigenmode analysis and computed sequences of low order f-modes.
Apart from them the traditional eigenmode problem of oscillations of rapidly rotating stars had been solved only within the Cowling approximation 
\citep{Yoshida_Eriguchi1997,*Yoshida_Eriguchi1999,*Yoshida_Eriguchi2001}, 
where Eulerian perturbations of metric coefficients are neglected.

Remarkable progresses in numerical relativistic hydrodynamics, however,
have made direct simulations of stellar oscillations possible. 
The Thessaloniki-MPA group computed oscillations in
the Cowling approximation\citep{Font_etal2001,*Stergioulas_etal2004}
and in the spatially conformally flat approximation\citep{Dimmelmeier2006}.
The AEI group computed non-linear
oscillations of rotating stars in full GR \citep{Baiotti_etal2007,*Manca_etal2007,
*Takami_etal2011}. The T\"{u}bingen group has developed 
linear evolution codes specialized for oscillations of rotating stars
with the Cowling approximation \citep{Kastaun2008,*Gaertig_Kokkotas2008,*Gaertig_Kokkotas2009,
*Krueger_etal2010,*Gaertig_Kokkotas2011,*Gaertig_etal2011}.
The T\"{u}bingen-LSU-Thessaloniki group computed the low order
f-mode sequences in full GR \citep{Zink_etal2010}.

Despite the remarkable successes in computing oscillations of relativistic rotating
stars by direct hydrodynamics simulations, the traditional eigenmode problem still needs to be studied. Firstly extracting eigenmodes by dynamical
simulations needs much more computational resources than the eigenmode
analysis. Especially the extraction of (the modulus of) an eigenfunction
needs multiple numerical runs to obtain a single mode \citep{Stergioulas_etal2004}.
Secondly it is not straightforward to identify eigenmodes by 
looking at the results of dynamical simulations. As the
observational asteroseismology \citep{Christensen-Dalsgaard}
the knowledge of the eigenmode analysis
is indispensable to interpret what is obtained by a numerical experiment.

In this paper we present a non-axisymmetric eigenmode analysis of rotating
relativistic stars beyond the Cowling approximation. 
For eigenmodes with low
order and degree, which may be mainly excited in an astrophysical
situation and are of interest in gravitational wave astronomy, 
the Cowling approximation works poorly with large errors 
\citep{Yoshida_Eriguchi1999,*Yoshida_Eriguchi2001}.
Thus we here take into account gravitational perturbation with suppressing
gravitational radiation. For typical compactness
of a neutron star ($\sim 0.2$), the damping time of a typical fluid oscillation
due to gravitational radiation is three orders of magnitude larger than the oscillation period 
(e.g., \citet{Andersson_Kokkotas1996}) and it may be neglected in evaluating
the characteristic oscillation frequencies. 

\section{Formulation}
In the context of relativistic astrophysics, approximations of GR
with the omission gravitational waves are considered 
by Isenberg (\citet{Isenberg2008}, originally written in 1978). 
The idea is that the behavior of a gravitating system may be well-approximated without gravitational emission as far as its timescale is longer than the dynamical one.
Isenberg proposed several recipes to eliminate the degree 
of gravitational wave from the theory.
Later \citet{WM1995} rediscovered
one of the approximations of \citet{Isenberg2008} and applied it to 
model quasi-equilibria of relativistic binaries (\citet{WMM1996,*Flanagan1999,*Wilson_Mathews_book}). Since then the
conformal flatness approximation (hereafter CFA ; also called 
IWM(Isenberg-Wilson-Mathews) theory or CFC(conformal- 
flatness condition)) has been adopted
in numerical relativity \citep{Cook_etal1996,*Oechslin_etal2002,*Dimmelmeier_etal2002,Dimmelmeier2006}.
The simplest introduction of CFA is done in (3+1)-decomposition of spacetime
 (see e.g., \citet{Gourgoulhon2012}). A general spacetime metric is written as
	 \footnote{We adopt the geometrized unit, $c=G=M_\odot=1$. The indicies
	 of tensors denoted by the upper alphabet ($a,b,c,\dots$) runs spacetime
	 indices $t,r,\theta,\varphi$, while the lower alphabet ($i,j,k,\dots$)
	 are restricted to the spatial indices $r,\theta,\varphi$.},
\begin{equation}
	ds^2 = -\alpha^2dt^2 + \psi^4\hatgam_{ij}(dx^i+\beta^idt)(dx^j+\beta^jdt).
\label{eq:3plus1metric}
\end{equation}
CFA assumes $\hatgam_{ij}$ to be a flat spatial metric $f_{ij}$. Assuming the trace
of the extrinsic curvature $K_{ij}$ of the spatial slice to be zero, we obtain the system
of elliptic partial differential equations \citep{Wilson_Mathews_book,*Gourgoulhon2012}:
\begin{equation}
	\hnab^i\hnab_i\psi = -2\pi\psi^5\left(\rho_{_H}
	+\frac{1}{64\pi\alpha^2}\hLam_{ij}\hLam^{ij}\right),
\label{H.C.}
\end{equation}
\begin{equation}
	\hnab^i\hnab_i(\alpha\psi) = 2\pi\alpha\psi^5
		\left(\rho h (3(\alpha u^t)^2-2)+5p
		+\frac{7}{64\pi\alpha^2}\hLam_{ij}\hLam^{ij}\right),
\label{alphapsi}
\end{equation}
\begin{equation}
	\hnab^i\hnab_i\beta^j = 16\pi\alpha\psi^4S^j-\frac{1}{3}\hnab^j(\hnab_k\beta^k) + \hLam^{jk}\hnab_k\ln\left(\frac{\alpha}{\psi^6}\right).
\label{M.C.}
\end{equation}
Here $\hnab_i$ means the covariant derivative with respect to
the flat metric $f_{ij}$. The indices of tensors are raised and lowered by $f_{ij}$.
$\hLam^{ij}$ is defined as $\hLam^{ij} = \hnab^i\beta^j+\hnab^j\beta^i-\frac{2}{3}f^{ij}\hnab_k\beta^k$. In the source terms we have
$\rho_{_H}\equiv n_an_bT^{ab}=\frac{\rho}{h}(\alpha\pi^t)^2+\alpha^2pg^{tt}$,
$S^j=-\delta^j_an_cT^a_c=\frac{\rho}{h}\alpha\pi^t\pi^j
 - \frac{1}{\alpha^2}\frac{\rho}{h}(\alpha\pi^t)^2$, where $n_a$ is the unit normal
 to the spatial hypersurface of $t=\mbox{const.}$.
 Linearized around the equilibrium state, these equations are elliptic PDEs for Eulerian perturbations of $\alpha$, $\psi$ and $\beta^j~(j=r,\theta,\varphi)$. 
For a given right hand side, these five equations are formally solved
for $\delta\alpha, \delta\psi, \delta\beta^j$, by using appropriate
Green's functions. For scalar variables $\delta\alpha$ and $\delta\psi$, we use
Green's function $1/|\vec{r}-\vec{r}'|$ which is expanded as a sum of 
scalar spherical harmonics. For the vector variable $\delta\beta^j$, we introduce
a vector harmonic expansion of the vector Laplacian operator as 
in \citet{Hill1954}. 

The assumptions on the equilibrium stars that we perturb are:
(1) stationary and axisymmetric, (2) the
equation of state (EOS) of the constituent fluid is barotropic.
For slowly rotating cases, we follow \citet{Yoshida_Kojima1997}
to construct the equilibria. In their treatment a stellar structure
is computed up to the first order of its rotational frequency,
i.e., physical quantities depend on the radial coordinate
and they coincide with those of the non-rotating ones
except for the $(t,\varphi)$-component
of metric coefficient \citep{Chandrasekhar_Miller1974}.
The EOS adopted in \citep{Yoshida_Kojima1997} is $p=K\epsilon^{1+1/n}$,
where $\epsilon$ is energy density, $K$ and $n$ are constants, and $p$ is pressure. For the comparison of slowly
rotating cases we use the same equilibria as \cite{Yoshida_Kojima1997}. For rapidly rotating cases
we use equilibria computed by COCAL code \citep{Uryu_Tsokaros2012}
assuming an EOS of polytropic form as $p=K\rho^{1+1/N}$,
where $\rho$ is rest mass density. 
It is known that the metric of a general stationary and axisymmetric 
spacetime cannot be cast into a conformally flat form on which we 
perform perturbation. Therefore, for internal consistency 
we use equilibria constructed by assuming conformal flatness,
\begin{equation}
	ds^2 = -\alpha^2dt^2 + \psi^4[dr^2+r^2d\theta^2
	+r^2\sin^2\theta(d\varphi-\omega dt)^2],
\label{eq:iwm equil}
\end{equation}
where $\alpha$, $\psi$ and $\omega$ are metric coefficients.

For simplicity we assume that perturbed fluid follows the same
EOS as in the equilibrium. This suppresses g-modes arising from
a stratified chemical composition and a beta-freezing of nucleons
in the perturbed fluid \citep{Reisenegger_Goldreich1992}. 
Thus we have
\begin{equation}
	\frac{\Delta p}{p} = \left(1+\frac{1}{n}\right)\frac{\Delta \epsilon}{\epsilon}
\end{equation}
for our slowly rotating cases and
\begin{equation}
	\frac{\Delta p}{p} = \left(1+\frac{1}{N}\right)\frac{\Delta\rho}{\rho}
\end{equation}
for our rapidly rotating cases. $\Delta$ here means Lagrangian perturbation
of the quantity following it. Relativistic enthalpy $h$ is defined as $h=(\epsilon+p)/\rho$. In the present case all the thermodynamic
variables $\epsilon, p,\rho$ are expressed as functions of $h$.

The perturbation equations for fluid variables are the equation
of rest mass conservation and the equation of motion (the conservation of
energy-momentum tensor). We introduce perturbed fluid variables $\delta h$
and $\delta\pi_a$, where $\delta$ means Eulerian perturbation of a quantity that follows it.
Momentum density $\pi_a$
is defined as $\pi_a = hu_a$, where $u_a$ is 4-velocity of the fluid.  Then
the perturbed rest mass conservation is written as
\begin{equation}
	\delta\left[\nabla_a\left(\frac{\rho}{h}\pi^a\right)\right] = 0.
	\label{eq:restmass_conserv}
\end{equation}
It should be noticed that $\nabla_a$ above is the covariant derivative
of the four dimensional metric. The perturbed equation of motion
is expressed as
\begin{equation}
	\delta\left[\pi^b\partial_b\pi_j-\pi^b\partial_j\pi_b\right] = 0
	\quad (j=r,\theta,\varphi)
	\label{eq:momentum_conserv}
\end{equation}

At a stellar surface where $p=0$, Eq.(\ref{eq:restmass_conserv}) is replaced by a boundary condition. We adopt the conventional condition
\begin{equation}
	\Delta p=0, 
	\label{eq:bc_surf}
\end{equation}
i.e., Lagrangian perturbation of pressure vanishes.
Lagrangian and Eulerian perturbations are related to each other
by using the Lie derivative 
along Lagrangian displacement vector which is a function of 
the perturbation of 4-velocity \citep{Friedman1978}.
Therefore the boundary condition is cast into
a relation between $\delta h$ and $\delta\pi_j$.

The scheme of solutions of eigenmode problem is a modified version
of \citet{Yoshida_Eriguchi1997}. The equations for fluid variables 
(Eq.(\ref{eq:restmass_conserv}), (\ref{eq:momentum_conserv}),(\ref{eq:bc_surf}))
are cast into a form in which all the fluid variables are on the left
hand side and the terms containing perturbed metric coefficients 
are on the right hand side. If the right hand
sides are omitted, 
the equations in the Cowling approximation used in \citet{Yoshida_Eriguchi1997}
are recovered.
Once the metric perturbation
is provided, the right hand sides are computed and the system of 
equations are solved for the fluid perturbation and the eigenfrequency.
On the other hand, Eqs.(\ref{H.C.}),(\ref{alphapsi}),(\ref{M.C.}) are inverted
when their source terms are computed for given fluid 
perturbations and the prescribed gravitational perturbations. Thus the 
equations for fluid and metric variables are iteratively solved.

\section{Results}
First we compare the eigenfrequencies of low-order non-axisymmetric modes
of slowly rotating stars obtained by the CFA, with those by the full GR theory.
In Table \ref{tabl:slowrot} eigenfrequencies computed by the CFA and the Cowling
approximation are compared with the real part of the quasi-normal frequencies
in full GR theory. $\sigma^0_{_{CFA}}$ and $\sigma^0_{_{Cw}}$ are the eigenfrequencies
of non-rotating stars for the CFA and the Cowling approximation, respectively.
$\sigma^0_{_R}$ is the real part of the quasi-normal frequency for a non-rotating
star. $\sigma'_{_{CFA}}, \sigma'_{_{Cw}},\sigma'_{_R}$ are rotational corrections
to the frequencies defined in \citep{Yoshida_Kojima1997},
\begin{equation}
	\sigma = \sigma^0 + m\sigma'\Omega,
	\label{eq:slowrot_mode}
\end{equation}
where $\Omega$ is the rotational frequency of the star.
It should be noted that although the metric of a spherical 
and static background spacetime can be exactly cast into a
conformally flat form (by introducing the isotropic coordinate), 
the perturbed spacetime around it is no longer exactly conformally
flat. The frequencies computed in the CFA are slightly different from the quasi-normal
ones with all the GR effects taken into account. As expected, the errors
in frequencies computed in the CFA are smaller than the corresponding ones by the Cowling. 
Although the errors in rotational correction (the factor of $\Omega$ in 
Eq.(\ref{eq:slowrot_mode})), $\sigma'$, seem slightly larger for the CFA compared to the Cowling, the errors in 
the frequencies for non-rotating case ($\sigma^0$) are much larger for Cowling. Thus the total error in the frequencies are smaller for the CFA. Overall, the error
in the frequencies of $m=2$ f-, $p_1$-, and $p_2$-modes for the CFA is within 1\% for the typical compactness of a neutron star. 

\begin{table}
\caption{Frequencies of the lowest quadrupole ($l=m=2$) modes for slowly
rotating stars.
The EOS is polytropic ($p=K{\rm \epsilon}^{1+1/n}$).
 $\sigma_{_R}$
is the real part of the quasi-normal frequency of a spherical star with given
compactness $M/R$ (taken from \citet{Yoshida_Kojima1997}).
$\sigma_{_{CFA}}$ is the eigenfrequency computed by the CFA and $\sigma_{_{Cw}}$
is the frequency obtained by the Cowling approximation.
$\sigma'_{_R}, \sigma'_{_{CFA}},\sigma'_{_{Cw}}$ are rotational corrections to
the corresponding frequencies (see text for the definition).
}
\begin{tabular}{cccccc}\hline\\
mode & $M/R$ & $\sigma^0_{_{CFA}}/\sigma^0_{_R}$ & $\sigma^0_{_{Cw}}/\sigma^0_{_R}$
& $\sigma'_{_{CFA}}/\sigma'_{_R}$ & $\sigma'_{_{Cw}}/\sigma'_{_R}$\\
\hline\\
f 	& 0.100 & 0.998 & 1.26 & 0.998 & 1.02\\
  	& 0.200 & 0.997 & 1.15 & 1.00 & 1.01\\
$p_1$  & 0.100 & 1.00 & 1.10 & 0.994 & 1.00\\
	& 0.200 & 0.997 & 1.11 & 0.991 & 1.01\\
$p_2$ & 0.100 & 1.000 & 1.05 & 0.997 & 1.00\\
	& 0.200 & 0.999 & 1.06 & 1.00 & 1.00\\\hline 
\end{tabular}
\label{tabl:slowrot}
\end{table}

Next we check errors of eigenmodes due to finite grid spacing. 
Since we work in the
finite-difference approximation of the equations, eigenmodes
obtained numerically depend on the grid spacing. By changing the grid
spacing, we measure the dependence of the error on it. When the finite grid parameter
$\delta$ is introduced (say, $\delta=\Delta r$ in the radial grid), a numerically-computed eigenfrequency is developed as a Taylor series around the
limit of $\delta\to 0$, thus $\sigma[\delta] = \sigma_0 + \sigma_1\delta + \sigma_2\delta^2 + \cdots$. Here $\sigma_0$ is the value of the
numerical eigenfrequency computed by Richardson extrapolation. To evaluate the
error size we compare the size of each
terms in the Taylor series. For $l=m=2$ f-mode of a non-rotating star with $M/R=0.200$, we have
 $\sigma_0=-1.17, \sigma_1=2.18, \mbox{and}~ \sigma_2=2.41\times 10^{-3}$, when $\delta$ is the radial grid spacing normalized by the radius of the equilibrium star. The rotational correction $\sigma'$ defined above is developed 
 in the same way and we obtain 
 $\sigma'_0=6.71\times 10^{-1}, \sigma'_1=4.36\times 10^{-1}, \sigma
'_2=-8.20\times 10^{-2}$.

Our code performs (vector) spherical-harmonic expansion of variables and equations with a truncation of series
at some value of the degree of the harmonics. For the $l=m$ f-mode,
a scalar quantity as $\delta h$ is expanded by 
$\{P_{m+2k}^m(\cos\theta)\}~(k=0,1,2,\dots,k_{\rm max})$ for a given
order $m$.
Thus the convergence 
in the angular grid is examined through the behavior of the eigenmodes
for different $k_{\rm max}$. For a rapidly rotating star of polytropic
index $N=1$ and $M/R=0.1$ with its surface
polar to equatorial axis ratio being 0.65, it is necessary to have $k_{\rm max}\ge14$ to determine an eigenfrequency within $1\%$ of error. 
A star closest to its mass-shedding limit at which the fluid 
on its equator rotates at Keplerian velocity, 
deforms more and the number of harmonic terms necessary to have
a good approximation of eigenfunctions increases.
For a star with a smaller rotational frequency a smaller number of $k_{\rm max}$ is sufficient.

For rapidly rotating cases, we compute sequences of the low-order counter-rotating f-modes of the order $m=2,3,4$. 
These modes correspond to $l=m$ f-modes in the non-rotating stars. 
The pattern speed of these modes are retrograde
with respect to the stellar rotation.
 They have been of great interest as because they are the most 
 susceptible modes to an instability driven by gravitational radiation (Chandrasekhar
-Friedman-Schutz (CFS) instability ; \citet{Chandrasekhar1970,Friedman_Schutz1978, Friedman1978}).
%
\begin{figure}
\includegraphics[scale=1.2]{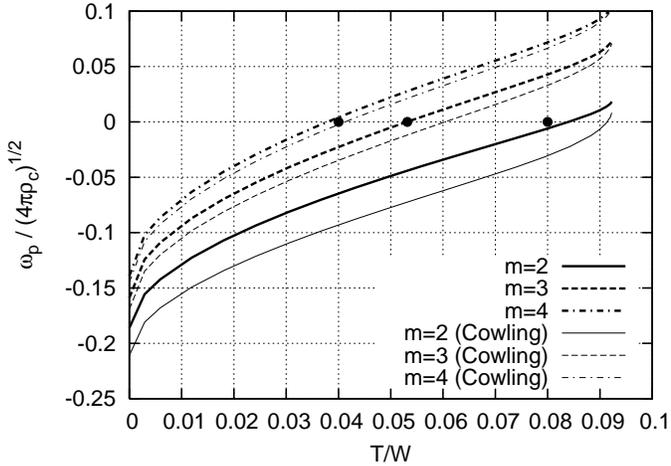}%
\caption{\label{fig:freqseq}
Dimensionless pattern speed $\omega_p$ (defined as $\sigma/m$ where $\sigma$
is the eigenfrequency and $m$ is the azimuthal quantum number of 
the eigenmode) of counter-rotating f-modes as a function of a stellar
rotational parameter $T/W$. These modes become $l=m$ f-modes
in the non-rotating limit of the star. For a slowly rotating star their
pattern rotates against the stellar spin. These modes becomes unstable
to the CFS instability at their zeroes when viscosity is neglected.
The rightmost points of the mode sequence correspond to the mass-shedding
limit of the star. The black dots are the neutral points of the instability
computed by SF98. From right to left they correspond to $m=2,3,4$ f-modes.
}
\end{figure}
%
\begin{figure}
\includegraphics[bb= 56 56 370 282, scale=0.8]{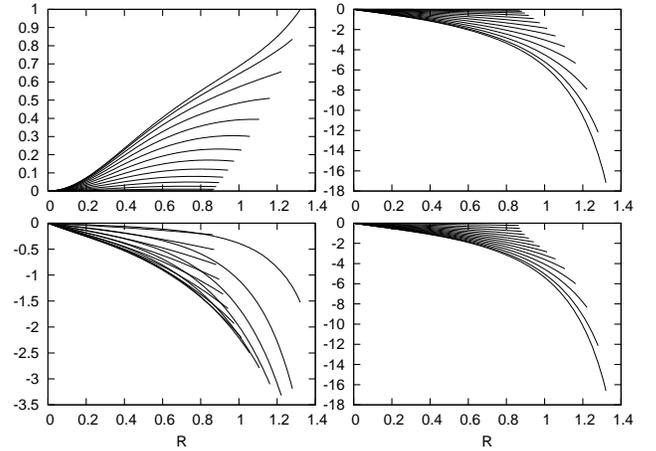}%
\caption{\label{fig:fluidpert}
Perturbed fluid variables of the counter-rotating bar mode near its neutral point of CFS instability
($T/W=0.0843$, the ratio of the polar to the equatorial radius is 0.650). The value of functions along each "$\theta=\mbox{const.}$" direction
is plotted as a function of $r$. 
Each panel corresponds to the following functions ; top-left:$\delta h$,
top-right:$\delta\pi_r$, bottom-left:$\delta\pi_\theta/r$, 
bottom-right: $\delta\pi_\varphi/r\sin\theta$.
}
\end{figure}
%
\begin{figure}
\includegraphics[bb= 56 56 370 282, scale=0.9]{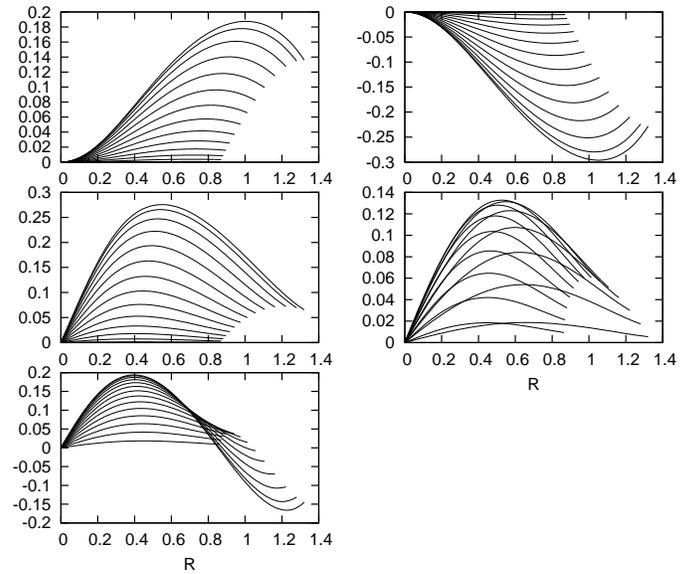}%
\caption{\label{fig:metricpert}
Metric perturbations for the same model as Fig.\ref{fig:fluidpert} are plotted inside the star.  
Each panel corresponds to the following functions ; top-left:$\delta\alpha$,
top-right:$\delta\psi$, centre-left:$\delta\beta_r$, centre-right $\delta\beta_\theta/r$ and bottom:$\delta\beta_\varphi/r\sin\theta$.
}
\end{figure}

In Fig.\ref{fig:freqseq} the pattern speed $\omega_p$ of counter-rotating f-modes with $m=2,3,4$
are plotted for the stellar equilibrium sequence of $N=1$ with a fixed
rest mass. It is defined as $\omega_p = \sigma/m$.
The ratio of the rotational energy to the gravitational
energy $T/W$ is adopted as a parameter characterizing the degree of stellar
rotation \citep{KEH1}.
The non-rotating limit of the equilibrium sequence have 
a compactness $C$ (i.e., the ratio
of gravitational mass to the circumferential radius)  value of 0.2.
When the gravitational radiation is taken into account, 
an inviscid star rotating faster than the zeroes of these modes
becomes unstable due to the CFS instability.
As is expected the difference between a mode sequence computed
in the CFA
and that computed in the Cowling approximation becomes smaller for larger $m$.
Introducing gravitational perturbations, the pattern speed tends to be 
larger than in the cases with the Cowling approximation. Thus the
Cowling approximation underestimate the CFS instability
\citep{Yoshida_Eriguchi1997}. It should be also remarked that
the slope of the pattern-speed curve around $T/W=0$
is larger than that around the actual neutral point. Therefore
an estimate of neutral points by using the slow rotation
approximation would rather overestimate the instability.

Eigenfunctions of perturbed fluid and metric variables are plotted
in Fig.\ref{fig:fluidpert} and \ref{fig:metricpert} respectively (although metric
variables are computed also in the vacuum domain, they are 
displayed inside the star). Each curve
corresponds to the value of a function on a "$\theta=\mbox{const.}$" line.
Notice that the ratio of the polar axis to
the equatorial axis of the star is $0.65$ and the rightmost point
of each curve corresponds to an equilibrium surface point.
These are plots for the counter-rotating bar mode at $T/W=0.0843$ where
it is neutrally stable against the bar-mode CFS instability. 
Eigenfunctions are normalized as $\delta h=1$
at the closest surface point to the equator. 
As a star spins up, the eigenfunctions in enthalpy and in the momentum change their profiles in such a way that their amplitudes are peaked more toward the equator \citep{Yoshida_Eriguchi1999}.

\section{Conclusion}
We presented a new code that solves the eigenmode problem
for rapidly rotating stars in GR. The code is intended to "go beyond
the Cowling approximation" on which the precedent eigenmode solvers 
in GR have been mainly developed. Our code adopts the CFA to take into account the non-radiative part of relativistic gravity. 
For slowly rotating stars we see that the mode frequency computed
by the CFA rather improves those obtained by Cowling approximation. 

A particularly interesting issue in rapidly rotating relativistic stars is their instability. The zeroes of counter-rotating modes marks the onset of the CFS instability and the slow rotation approximation is rather erroneous in 
evaluating them. Thus the application of the current code on this issue is promising. 
We compared the critical $T/W$ available through Fig.5
in SF98 by reading off the numbers there and
compared them with our results. For the weaker gravity $C=0.1$ sequence,
the difference of the critical $T/W$ for the CFA from that of SF98 is 
less than 1\% for $m=3$ and 2\% for $m=4$ mode. 
For $C=0.2$ the difference is 6\% for $m=2$ and for $m=4$
and less than 1\% for $m=3$. The good agreements for $m=3$ might be
coincidental. These results are consistent with the analysis done by
non-linear simulations\citep{Zink_etal2010}.


\begin{acknowledgments}
The author thanks K\={o}ji Ury\={u} for providing the numerical data of
equilibrium rotating stars computed with the COCAL code and his useful comments. He also thanks Luca Baiotti for carefully reading the manuscript
and Yoshiharu Eriguchi for his useful comments.
\end{acknowledgments}

\bibliography{cfa_ref}

\end{document}